\documentclass[12pt]{article}
\usepackage{graphicx}

\begin{document}

\title{Couplings of $N N (n \pi)$ $n \geq 1$ }

\author{Bing An Li \\
Department of Physics and Astronomy, University of Kentucky\\ Lexington, Kentucky,
40506, USA}
\maketitle

\newpage

\pagebreak
\begin{abstract}
Determinations of the couplings of $N N (n \pi) (n \geq 1)$ are reported.
The study is based on both a quark model of nucleon  and a chiral field theory of mesons.
The coupling of $N N \pi$ is predicted and is in agreement with current value.  
It shows that the coupling $ N N 2 \pi $ is resulted in the nature that pion is a Goldstone boson. 
The couplings
of $N N (n \pi) (n \geq 2)$ are predicted by this approach.  

\end{abstract}
\newpage

It is known that the  $N N \pi$ vertex is an important coupling of the strong interactions. 
This coupling is a problem of nonperturbative QCD.
There are already extensive studies on this coupling by various approaches. 
The value of the coupling constant 
of nucleon and pion has been determined [1].
Besides the$ N N \pi$ coupling the $N N 2\pi$ vertices have been introduced to study the $\pi-N$ s-wave 
interaction in Ref. [2] and these $2 \pi$ couplings [2] have been used to study the N - N potential in Ref. [3].
In Refs. [4] the current algebra and the PCAC have been applied to calculate the matrix elements in which additional 
soft pions are emitted or absorbed.

In this paper a quark model of nucleon [5] and a chiral field theory of mesons [6] are employed to determine the 
N-$\pi$ coupling and to explore the nature of chiral symmetry of the couplings of N-2$\pi$ and $N N (n \pi) (n \geq  3)$.

In Ref.[5] we have proposed  an approach of studying the electromagnetic and the weak interactions of nucleon (baryon) in 70's.
In this approach SU(6) symmetry is applied to construct the quark wave functions of 
baryons in the rest frame and the transition matrix elements. The effective electromagnetic and the charged weak 
currents of quarks are constructed and applied to study the electromagnetic and the charged weak processes of baryons. 
In this approach the $G^p_M(q^2)$, 
a parameter from the model, and the $G^A(0)$ are taken as inputs. 
In 1975 this model predicted that
in small range of $q^2$ the ratio $\mu_p\;G^p_E(q^2)/G^p_M(q^2)$ is about $\simeq 1$ and decreases with $q^2$ when $q^2 > 1\; \textrm{GeV}^2  $. 
These predictions agree with the new data. This model also predicts a very small $G^n_E(q^2)$ and
$G^n_M =- {2\over3} G^p_M$. The $\mu_{p\rightarrow \Delta} = 1.23\; {2\sqrt{2}\over 3} \mu_p$, 
the helicity amplitudes of $\gamma\; p\rightarrow \Delta$, and very small negative $S_{1+}$, $E_{1+}$ are predicted too. By the way in this 
model the shape of a proton is spherical in the rest frame. In the moving frame the shape of the proton is no longer
spherical because of Lorentz contraction.
For weak interactions of the charged currents the axial-vector form factor $G^A(q^2)/G_A(0)$ is predicted and in very good agreement with the data. 
The predicted cross sections of the scattering of the neutrino (antineutrino) and the nucleon by the charged 
weak currents are in good agreements with data too.

This quark model of nucleon is applied to study the strong interactions of nucleon - pions in this paper.
It is very important to notice that the pion is a Goldstone boson and it is massless in the chiral limit.
The couplings between quarks and pions are obtained from a chiral field theory of mesons 
that we have proposed in Ref. [6].
In this chiral field theory the pion is a Goldstone boson and the mass of the pion 
is resulted in explicit chiral symmetry breaking by the current quark mass. 
The pion mass [6,7], the pion form factor [8], $\pi-\pi$ scattering [6], and the anomaly $\pi^0\rightarrow \gamma\gamma$ [6,9] etc.
are studied and the theory agrees with data well. 
The physics processes of the pseudoscalar, the vector, and the axial-vector mesons are studied too. Theory is in good agreement with data.

The nucleon model and the chiral field theory of mesons are applied to study the strong interactions of nucleon and pions
in this paper. For convenience 
the Lagrangian of the chiral field theory of mesons in the case of two flavors Eq. (1) of Ref. [6] is copied as following
\begin{eqnarray}
{\cal L}=\bar{\psi}(x)(i\gamma\cdot\partial+\gamma\cdot v
+\gamma\cdot a\gamma_{5}
-mu(x))\psi(x)\nonumber \\
+{1\over 2}m^{2}_{0}(\rho^{\mu}_{i}\rho_{\mu i}+
\omega^{\mu}\omega_{\mu}+a^{\mu}_{i}a_{\mu i}+f^{\mu}f_{\mu}),
\end{eqnarray}
where $\psi$ is the quark fields, \(a_{\mu}=\tau_{i}a^{i}_{\mu}+f_{\mu}\), \(v_{\mu}=\tau_{i}
\rho^{i}_{\mu}+\omega_{\mu}\), 
\(u=e^{i\gamma_{5}(\tau_{i}\pi_{i}+
\eta)}\), the quantity $m$ is originated in the quark condensate, and the $m_0$ is related to the mass of the $\rho$ meson [6].
The kinetic terms of the meson fields are generated from the quark loop diagrams
and the physical meson fields are defined [6]. For example 
\[\pi^i \rightarrow {2\over f_\pi} \pi^i\;\; (physical) \]
where the pion decay constant $f_\pi$ is determined 
by the kinetic term of the pion field and $f_\pi = 0.186\;\textrm{GeV}$ is taken.
Another parameter g is generated from the kinetic term of the $\rho$ field
and $g=0.395$ is determined by 
input the decay rate of $\rho\rightarrow e e^+$. The m is determined by $f_\pi$ and g to be
\(m = 0.242\; \textrm{GeV}$.
 
There are two sources of the pion field:
the non-linear $\sigma$ term in Eq. (1) is the first source and
the mixing between the pion and the axial-vector fields [6] generates another pion field
\[-{c\over g}{2\over f_\pi} \partial_\mu \pi^i,\]
where \({c\over g} = \frac{f^2_\pi}{2g^2 m^2_\rho}\). 
The couplings between the quark fields and the pion fields is found from Eq. (1)
\begin{equation}
{\cal L} = - \bar{\psi} \{{2im\over f_\pi}\gamma_5 \tau^i \pi^i + {c\over g}{2\over f_\pi} \gamma_\mu \gamma_5 
\tau^i \partial_\mu \pi^i\} \psi.
\end{equation}
In the model of nucleon [5] the effective EM current of the quarks has a term of anomalous magnetic moment 
of the quark and the
effective charged weak current of the quarks is expressed as
\[\lambda \gamma_\mu \gamma_5\]
where the parameter $\lambda$ is determined as
\[\lambda = {5\over3} G_A\] 
which is the same as the one obtained in the model of SU(6) symmetry. In Eq. (2) the axial-vector part \(\tau^i \gamma_\mu \gamma_5\)
is the same as the charged weak current. Therefore, the effective Lagrangian of Eq. (2) should be rewritten as
\begin{equation}
{\cal L} = - \bar{\psi} \{{2im\over f_\pi}\gamma_5 \tau^i \pi^i + {c\over g}{2\over f_\pi} \lambda \gamma_\mu \gamma_5
\tau^i \partial_\mu \pi^i\} \psi.
\end{equation}
Eq. (3) is used to calculate the transit matrix element of $N\rightarrow N + \pi$ 
and to determine the coupling of $i \bar{N} \gamma_5 \tau^i N \pi^i$.
Before the calculation it is 
worth to mention that this theory has both the features of the current algebra and the non-perturbative QCD.
In Refs. [4] the current algebra and PCAC have been used to study multi-pion production and absorption.
Based on the vector currents, the axial-vector currents, and the non-linear $\sigma$ model of the current algebra 
the vector mesons, the axial-vector mesons and the pseudoscalar mesons are introduced to this Lagrangian (1) 
which is $U(2)_L\times U(2)_R$ global chiral symmetric.
Some of the results of current algebra are revealed from this theory [6], for example, 
the scattering lengths and the slopes of the $\pi - \pi$ scattering and the two Weinberg sum rules [10].
The PCAC is satisfied [11].
On the other hand, this chiral field theory has some of the major features of nonperturbative QCD: 
1) the quark loop diagrams are at the leading order in the $N_C$ expansion and the meson loop diagrams are
at higher orders; 2) this theory has the explicit chiral symmetry in the limit $m_{light\; quarks}\rightarrow 0 $
and it is broken explicitly by the current quark masses;  
3) there is dynamical chiral symmetry  broken by the quark condensation which plays a very important role
in this theory.  

Replacing the EM currents or the charged weak currents in the transit matrix elements of the nucleon Eqs. (49, 197) in Ref. [5R]
by the vertex of quark - pion (3),
the $\pi N$ coupling is obtained by calculating the transit matrix element $N \rightarrow N + \pi$ 
\begin{eqnarray}
<B_\lambda(p')_{U^{\prime}} \pi(q) | S |B_\lambda(p)_U> = i(2\pi)^4 \delta (p - q - p') 
{1\over2}\int dx'_1 dx'_2 dx_1 dx_2 \nonumber \\
\bar{B}^{\lambda^{\prime}}
_{\alpha\beta\gamma,ijk_1}(x'_1,x'_2,0)_{U^{\prime}}  (\tau^i)_{k_1 k'_1} \{-{2im\over f_\pi}\gamma_5 \pi^i 
- {c\over g}{2\over f_\pi}  \lambda \gamma_\mu \gamma_5 \partial_\mu \pi^i\}_{\gamma\;\gamma'} \nonumber \\  
M(x'_1,x'_2,x_1,x_2) 
B^\lambda_{k'_1 j i,\gamma'\beta\alpha}(0, x_2, x_1)_U,
\end{eqnarray}
where the B with indices are the nucleon wave functions Eqs. (33, 46) in Ref. [5R], 
\(U^{\prime}\) and U are the flavor states of the final and initial baryons [5] respectively, 
the function $M(x'_1,x'_2,x_1,x_2)$ defined in Ref. [5] is the effect of other two quark lines in the transit matrix element
and by the requirement of the SU(6) symmetry it is simply assumed that $M(x'_1,x'_2,x_1,x_2)$ is a scalar function 
and it is unknown, which is absorbed by the form factors defined in Ref. [5].
It is assumed that in the rest frame the wave functions contain s-wave only and they observe SU(6) symmetry. Using Lorentz transformation 
the wave functions are boosted to moving frame. 
The details can be found in Ref. [5]. 
There are two Lorentz invariant spacial functions $f_{1,2}(x_1, x_2, x_3)$ in the quark wave functions of nucleon and the assumption
\(f_2 = a f_1 \) is made [5] and \(a = 4.51 \) is determined. It is shown that this assumption works well in the range of $q^2 < 5\; \textrm{GeV}^2$.

Calculating the matrix element (4),
it is obtained
\begin{eqnarray}
{\cal L}_{NN\pi} = g_{N N \pi} F_{ N N \pi}(q^2) i \bar{N} \gamma_5 \tau^i N \pi^i,\\
g_{N N \pi} = {2 m_N G_A\over f_\pi} \{\frac{m}{m_N G_A} (\frac{1}{a} + {5\over 3} a - 1) - {2c\over g}\},\\
F_{ N N \pi}(q^2) = D_2(q^2) \{1 - \frac{1}{12 f_\pi} {1\over g_{ N N} \pi} \{10 m a -
{2c\over g}\lambda (6 a +4) m_N \} \frac{q^2}{m^2_N}\},
\end{eqnarray}
where q is the momentum of the pion, $q^2 = (p -p')^2$, and
\[D_2(q^2) = {1\over \mu_p} G^p_M(q^2)\frac{1}{ 1 + 2.39 {q^2\over 4m^2_N}} = \frac{1}{(1 + {q^2\over 0.71})^2 (1 + 2.39 {q^2\over 4m^2_N})}\]
Eq. (112) in Ref. [5R]
determined by fitting the magnetic form factor of the proton.

The coupling constant $g_{N N \pi}$ is defined at $q^2 = 0$ of the form factor of the $N N \pi$ vertex. 
If taking \(q^2 = m^2_\pi\), there is a very small correction from Eq. (6). 
The $m_N$ and the $G_A$ are known and all other the parameters in Eqs. (6,7) are determined in Ref. [5,6].
It obtains 
\begin{equation}
\frac{g^2_{N N\pi}}{4\pi} = 13.05.
\end{equation}
This value is consistent with the values of the $g_{ N N \pi}$ presented in Ref. [1]. 
It is worth to point out that because 
\[\frac{m}{m_N G_A} \{{1\over a} + {5\over 3} a - 1\}  - {2c\over g} = 1.01 \]
the Goldberg-Treiman relation is satisfied pretty well in this approach. 
The satisfaction of the Goldberg-Treiman relation means the parameters determined 
from the EM and the weak interactions of the nucleon in Ref. [5] and the meson theory [6] are reasonable.
Eq. (6) shows that there is cancellation between both the pseudoscalar and the axial-vector parts 
of the coupling constant $g_{N N \pi}$. 
Inputting the values of the parameters the form factor $F_{ N N \pi}$ is determined as
\begin{eqnarray}
F_{N N \pi}(q^2)  = D_2(q^2) \{ 1- 0.096 {q^2\over m^2_N}\}.
\end{eqnarray}
The radius of this form factor is revealed from Eq. (9)  
\begin{equation}
< r > = 0.916\; fm. 
\end{equation}
Eq. (9) shows that the form factor of the nucleon-pion decreases faster with $q^2$ than ${1\over \mu_p}G^p_M(q^2)$
whose radius is $0.81\; fm$.
The charged axial-vector form factor of nucleon
${1\over G_A(0)}G_A(q^2) = D_2(q^2) (1 + 1.125 {q^2\over M^2_N})$ Eq. (218) in Ref. [5R] 
agrees with data well and its radius is $0.72 \;fm$. $F_{N N \pi}(q^2)$ deceases faster than ${1\over G_A(0)}G_A(q^2)$ too.

In Refs.[2,3] the vertices between nucleon and two pions are introduced. 
Besides the coupling $NN\pi$ why it is needed to introduce the vertex $NN\pi\pi$ ?
In this approach this question is answered.
In the chiral limit $m_q\rightarrow 0$ Lagrangian (1) is $U(2)_L\times U(2)_R$ 
chiral symmetric and the pion is massless. By adding the quark mass matrix $-\bar{\psi} M \psi$ to the Lagrangian (1) [6,7],
where M is the mass matrix of the u and d quarks, the chiral symmetry is explicitly broken by current quark mass. 
In Eq. (1) the u matrix is via the nonlinear $\sigma$ model of the current algebra introduced to make the Lagrangian (1) chiral symmetric.
If taking the pion field into account only
\begin{equation}
u = e^{i\gamma_5 \tau^i \pi^i} = 1 + i \gamma_5 \tau^i \pi^i - {1\over2} \pi^i \pi^i + ...
\end{equation}
Eq. (11) shows that besides the coupling of quarks and one pion there are coupling between the quarks and two pions
\begin{equation}
-im \bar{\psi}\gamma_5 \tau^i \psi \pi^i + {m\over 2} \bar{\psi} \psi \pi^2.
\end{equation}
Using Eq. (12), the pion mass in the order of $O(m_q)$ is revealed from the calculations of two quark loops [6,7]
\begin{equation}
m^{2}_{\pi} = -{1\over3} {4\over f^{2}_{\pi}}
<\bar{\psi}\psi>(m_{u}+m_{d}).
\end{equation}
where the quark condensate is from the m (1). This is the Gell-Mann, Oaks and Renner formula [12].
Eq. (13) shows that in the chiral limit, $m_q = 0$, the pion is massless and a Goldstone meson.
After adding the quark masses, there are two mass dimensions m and the current quark mass.
Why the $m^2_\pi\propto m_{u} + m_{d}$ not $m^2_\pi\propto m$ ?
The reason is that there are cancellations between the two quark loops of the two vertices (12).     
If only one of the two terms of Eq. (12) 
is taken into account \(m^2_\pi \propto m \) is revealed.
Therefore, besides the coupling of 
\[-im \bar{\psi}\gamma_5 \tau^i \psi \pi^i\]
the vertex of two $\pi$ 
\begin{equation}
{m\over 2} \bar{\psi} \psi \pi^2
\end{equation} 
is necessary for pion to be a Goldstone boson. In this approach the vertex (14)
leads to the $N N \pi\pi$ coupling. Therefore, the $N N \pi\pi$ coupling is a necessary vertex and 
is resulted
in that the pion is a Goldstone boson.

Replacing the quark vertex of one pion (3) in Eq. (4) by the vertex of two pions (14), the form factor and coupling constant of the 
the nucleon - two pions vertex are determined as
\begin{eqnarray}
{\cal L}_{NN 2\pi} = F_{NN2\pi} (q^2) {1\over f_\pi} \bar{N} N \pi^2,\\
F_{NN2\pi}(q^2) =  {6m\over f_\pi} D_2(q^2) \{{1\over a} + a -1 - {a\; q^2\over 4 m^2_N}\},
\end{eqnarray}
where q is the total momenta of the two pions and \(q^2 = (p' - p)^2\). 
The coupling constant is defined as
\begin{eqnarray}
g_{NN2\pi} = F_{NN2\pi}(0) =
{6 m\over f_\pi} ({1\over a} + a - 1) = 29.13.
\end{eqnarray}
It is worth to point out that in this approach there is no contact interaction like
\[\bar{N}\vec{\tau}\gamma_\mu N \cdot \vec{\pi}\times \partial_\mu \vec{\pi}\]
introduced in Refs. [2,3]. However, in this approach this term can be obtained from the two couplings:
the nucleon is coupled to the $\rho$ meson and the $\rho$ couples to two pions and it is not a contact term.  
In this paper only
contact interactions between nucleon and pions are studied. On the other hand, the value of the coupling 
constant ${1\over f_\pi} g_{NN2\pi}$ is much greater than the corresponding constant presented in Refs. [2,3].

In the same way 
the coupling constants and the form factors between nucleon and multi-pions, $N N (n \pi)$ with $n > 2$ 
can be predicted too. 
Taking the pion into account only, 
the interactions between quarks and pions are expressed as
\begin{eqnarray}
- m \bar{\psi}  e^{i \gamma_5 \pi} \psi = -i m {2\over f_\pi} \bar{\psi} \gamma_5 \pi  \psi  
- i {2m\over f_\pi} \bar{\psi}\gamma_5 \tau^i \psi \pi^i \sum^{\infty}_{n=1} \frac{1}{(2 n +1) !!}(-1)^n ({4\over f_\pi^2} \pi^2)^n \nonumber \\
+ \bar{\psi} \psi {4m \over f^2_\pi} \pi^2 \sum^{\infty}_{n=1} \frac{1}{(2n)!!} (-1)^{n} ({4\over f^2_\pi} \pi^2)^{n-1}. 
\end{eqnarray}
As studied above, the shifting of the $a_\mu$ field is another source of the pion field. 
Using the couplings between quarks and 
odd number of pions with $n \geq 3$ and the procedure showing above, 
the couplings between nucleon and pions with odd number are determined as
\begin{eqnarray}
{\cal L}_{odd\; \pi} = F_{odd\; \pi}(q^2) i \bar{N} \gamma_5 \tau^i \pi^i N 
\sum^{\infty}_{n=1}\frac{1}{(2n+1)!!} (-1)^n ({4\over f^2_{\pi}} \pi^2)^n,\\
F_{odd\; \pi}(q^2) = g_{odd} D_2(q^2)\{1 - {5\over 6}\frac{m a}{ f_\pi}{1\over g_{odd}} {q^2\over m^2_N}\},\nonumber \\
g_{odd} = {2m\over f_\pi} ( {1\over a} +{5\over 3} a -1) = 17.53,\nonumber \\
F_{odd\; \pi}(q^2) = g_{odd} D_2(q^2) \{1 - 0.317\; q^2\},
\end{eqnarray}
where q is the total momenta of the pions.
 
Using the couplings between quarks and pions with even numbers ($ \geq 2$) (18) and the transit matrix element (4)
the couplings between nucleon and pions with even numbers are determined
\begin{eqnarray}
{\cal L}_{even\; \pi} = F_{even\; \pi}(q^2) {2\over f_\pi} \pi^2 \bar{N} N \sum^{\infty}_{n=1} \frac{1}{(2n)!!} (-1)^{n} ({4\over f^2_\pi} \pi^2)^{n-1}, 
\end{eqnarray}
where $F_{even\; \pi}(q^2)$ is the same as $F_{NN2\pi}(q^2)$ (16) and $g_{even} = F_{even\; \pi}(0) = g_{NN2\pi}$ (17).
The form factor is expressed as
\begin{equation}
F_{even\; \pi}(q^2) = g_{even} D_2(q^2) \{ 1- 0.343 q^2\}.
\end{equation}
  
It is straight forward to extend the study above to the cases of three flavors.
The chiral field theory of mesons of three flavors have been proposed [13] and the
physics of the axial-vector, the vector, and the pseudoscalar mesons of three flavors are studied.
Theory agrees with data well. 
The flavor wave functions of baryons can be found in Ref. [5]. 
In the chiral limit, the couplings between 
baryons and all the pseudoscalar mesons (pion, kaon, $\eta$, and $\eta'$) can be determined 
without new parameter. Using the Lagrangian (1), in the same way the couplings between the baryons 
and the vector and the axial-vector mesons can be determined too.
The detailed study is beyond the scope of this paper. 
 
In summary, in this paper a quark model of baryon and a chiral field theory are applied to study the couplings between 
nucleon and pions. The coupling constant $g_{NN\pi}$ predicted is consistent with current value and 
the nature of Goldstone boson of pion leads to the existence of the  $NN2\pi$. It predicted the existence of 
the couplings of nucleon and multi-pions. The predictions made in this paper can be tested experimentally.

\end{document}